\documentclass[a4paper,11pt]{article}
\usepackage[margin=23mm]{geometry}
\usepackage[american]{babel}
\usepackage[utf8]{inputenc}
\usepackage[T1]{fontenc}

\usepackage{lmodern}
\usepackage{csquotes}
\usepackage{microtype}
\usepackage{tikz}
\usepackage{amsmath,amssymb,amsthm,stmaryrd}
\usepackage{mathtools}
\usepackage{thmtools} 
\usepackage{todonotes}
\usepackage{biblatex}
\usepackage[hidelinks]{hyperref}
\usepackage[capitalize,nameinlink]{cleveref}

\newcommand{\Iv}[1]{\llbracket #1\rrbracket}
\newtheorem{theorem}{Theorem}
\newtheorem{lemma}[theorem]{Lemma}
\newtheorem{corollary}[theorem]{Corollary}
\providecommand\with{}
\newcommand\SetSymbol[1][]{\nonscript\,#1\vert \allowbreak \nonscript\,\mathopen{}}
\DeclarePairedDelimiterX\set[1]{\lbrace}{\rbrace}{ \renewcommand\with{\SetSymbol[\delimsize]} #1 }

\renewcommand*{\S}{\mathcal S}

\title{Counting $6$-Cliques in $O(n^{5.97})$ time and $O(\log n)$ space\thanks{All authors are supported by the project COALESCE (``Combinatorially and algebraically decomposing search spaces'') that has received funding from the European Research Council (ERC), grant agreement No 101171685.}}
\usepackage[affil-sl]{authblk}
\author{Afrouz Jabal Ameli}
\author{Niklas Mohrin}
\author{Jesper Nederlof}
\affil{Utrecht University}

\makeatletter
\patchcmd{\@maketitle}
  {\vskip 1em{\large \@date}}
  {}
  {}
  {\PackageWarning{nodate}{Could not remove the date block}}
\makeatother
\begin{document}
\maketitle
\begin{abstract}
	We present an algorithm that, given read-only access to the adjacency matrix of an $n$-vertex graph $G$, computes the number of $6$-cliques in $G$ using $O(n^{5.97})$ time and $O(\log n)$ space.
	We achieve this by exploiting a decomposition of an equality indicator function underlying, among others, the refutations of the Exponential Time Hypothesis by OpenAI and Anthropic, and of the $k$-Clique Conjecture by Vassilevska Williams and Zamir~[arXiv, 2026].
	Via known reductions, this implies an~$O(1.994^n)$ time and polynomial space algorithm for \textsc{Max Cut} on $n$-vertex graphs.
\end{abstract}
\section{Introduction}
A \emph{clique} is a set of pairwise adjacent vertices.
Counting fixed-size cliques is a basic problem in graph algorithms. Trivial enumeration counts all $k$-cliques in $n^{k}$ time.
For decades, the fastest known algorithms were based on fast matrix multiplication.
The classic reduction of Ne\v{s}et\v{r}il and Poljak~\cite{NP85} counts $k$-cliques in $n^{\omega k/3+o(1)}$ time when $k$ is divisible by three, where $\omega$ is the matrix multiplication exponent; rectangular matrix multiplication gives refined bounds for other values of $k$~\cite{EG04}.
Until recently, this was the only known route to a polynomial saving over trivial enumeration.

Recent developments go beyond this standard reduction.
Vassilevska Williams and Zamir~\cite{VWZ26} count $k$-cliques in $n^{(\omega/3-c)k}$ time for all sufficiently large fixed $k$, for an absolute constant $c>0$ independent of both $k$ and $\omega$.
The key step in each of these mentioned recent developments are various algebraic decompositions for the equality indicator function that also underlie the recent refutations of the Exponential Time Hypothesis~\cite{Anthropic26,OpenAI26}.

Given read-only random access to the adjacency matrix, trivial enumeration uses only $O(\log n)$ space for every fixed $k$.
Despite the progress on running time, no $O(n^{k-\varepsilon})$-time algorithm using only $(\log n)^{O(1)}$ space was previously known for any fixed $k\geq 3$ and constant $\varepsilon>0$.
The matrix multiplication algorithms do not directly yield such a result.
Their standard implementations that run in time linear in the tensor rank of the matrix multiplication tensor achieve their time bounds by sharing intermediate computations between many linear forms and storing polynomial-size arrays.
Evaluating the linear forms separately avoids these arrays, but its cost depends on the \emph{total support} of the decomposition rather than its rank: for known decompositions, this direct evaluation takes more than $n^k$ time.

\paragraph{6-Clique in Logarithmic Space.} Building on the mentioned recent developments, we present a decomposition with small enough total support (i.e.\ a \emph{sparse} decomposition) that achieves a polynomial saving in the running time and logarithmic space:
\begin{theorem}
	\label{thm:clique}
	The number of $6$-cliques in an $n$-vertex graph can be computed in
	$O(n^{5.97})$ time and $O(\log n)$ space, given
	read-only random access to its adjacency matrix.
\end{theorem}
By the standard reduction of~\cite[Section 2.2]{NP85}, Theorem~\ref{thm:clique} can also be used to count (induced) pattern occurrences of any $6$-vertex pattern graph.
The time--space complexity of fixed-size clique detection has received
attention in earlier work.
Woeginger~\cite[Open Problem~4.3(a)]{Woeginger08} asked for algorithms
with improved time and space complexities.
Vassilevska~\cite{Vassilevska09} gave a combinatorial algorithm that,
for every fixed $k\geq 3$ and $\alpha>0$, detects a $k$-clique in
$O(n^k/(\alpha\log n)^{k-1})$ time using $O(n^\alpha)$ space.
The stronger goal of obtaining $n^{k-\varepsilon}$ time and
polylogarithmic space for some fixed $k\geq 3$ and $\varepsilon>0$
was explicitly raised by Williams~\cite[Open Problems]{Williams16}.

\paragraph{Our Approach.} Like all of these previous recent developments~\cite{ALPPW26,CIMY26,Anthropic26,Fischer26,OpenAI26,VWZ26}
our key step is a decomposition of an equality indicator function. We use it to check equality on $4$ vertices $\{a,b,c,d\}$ and their copies $\{a',b',c',d'\}$ of the $6$-clique $\{a,b,c,d,e,f\}$.
Viewing vertex pairs as strings, the decomposition expresses the equality indicator $\Iv{ab=a'b',cd=c'd'}$ as a sum of terms indexed by a set of swapped indices.
At these indices, $ab$ is matched to $c'd'$ and $cd$ to $a'b'$; correction factors (which we later factor using \emph{anti-commutative indicators}) on the remaining indices make the identity exact.
We provide two complementary ways to evaluate each resulting term (see also Figure~\ref{fig:edge-test-split}).
\begin{itemize}
\item The \emph{factored evaluation} exploits a sparse factorization to separate the clique-counting expression into a product of sums: after fixing $e$, one factor checks whether $\{a,b,c,d\}$ is a $4$-clique and whether $ae,be\in E(G)$, while the other splits into two independent sums over pairs of vertices once $f$ is fixed as well.
This evaluation is efficient when few indices are swapped, because the correction factors admit a sparse factorization.
\item The \emph{direct evaluation} leaves the equality term unfactored and regroups the edge tests to separate the sum over $e$.
It benefits from having many swapped indices: the resulting equalities between $(a,b)$ and $(c',d')$ restrict the assignments for which this sum must be computed.
\end{itemize}

Choosing the faster evaluation for each term yields savings that outweigh the overhead of the decomposition, giving a polynomial improvement over naive enumeration while using only logarithmic space.
As mentioned before, our methods are similar to the recent methods for \textsc{Hyperclique}~\cite{ALPPW26,CIMY26,Fischer26}, and most closely related to the one for \textsc{Clique}~\cite{VWZ26}. In comparison with~\cite{VWZ26}, we choose between sparse factored evaluation and direct evaluation for each term to obtain a logarithmic-space algorithm, whereas~\cite{VWZ26} combines matrix multiplication with recursive reductions to sparser patterns.

\paragraph{Application to Max Cut.} In the \textsc{Max Cut} problem one is given an $n$-vertex undirected graph $G$ and asked for a subset $X$ that maximizes $|\set{ \{v,w\} \in E(G) \with v \in X,w \notin X }|$. The problem can be solved in $O(2^{\omega n/3})$ time and exponential space by a reduction to \textsc{$3$-Clique} by Williams~\cite[Section~3]{Williams05}.
Using his reduction with six parts reduces unweighted \textsc{Max Cut} to
polynomially many $6$-clique counting instances on
$O(2^{n/6})$ vertices. We represent these instances
implicitly, answering each adjacency query in polynomial
time and space. Then Theorem~\ref{thm:clique} implies:

\begin{corollary}
	\label{cor:maxcut}
	\textsc{Max Cut} on $n$-vertex graphs can be solved in
	$O(1.994^n)$ time and polynomial space.
\end{corollary}

This resolves the qualitative form of Woeginger's question~\cite[Open Problem~4.7(b)]{Woeginger08} by giving a polynomial-space algorithm with running time $O((2-\varepsilon)^n)$ for some constant $\varepsilon>0$.\footnote{Woeginger states the numerical target $O(1.99^n)$; our bound establishes a base strictly smaller than $2$, but does not reach that target.}

\paragraph{Notation.}
Throughout this paper we fix an undirected simple input graph $G$ and let $V(G)$ and $E(G)$ denote its vertex and edge set, respectively. We abbreviate $n = |V(G)|$ and $[\ell]=\{1,\ldots,\ell\}$. Write~$\Iv{P}$ for the indicator of a statement $P$, that is, $\Iv{P} = 1$ if $P$ is true, and $0$ otherwise. We let $\widetilde O (\cdot)$ omit factors polylogarithmic in $n$.

\section{Decomposing String Equalities}
\label{sec:string-identities}
We first collect identities about equality indicators, similar to other recent works~\cite{ALPPW26,CIMY26,Anthropic26,Fischer26,OpenAI26,VWZ26}. Let $q \geq 2$ denote the alphabet size, which we assume to be constant. 
For strings $s, s', w, w' \in [q]^\ell$, we decompose the equality indicator function $\Iv{s = s', w = w'}$ as follows:
\begin{align*}
	\Iv{s = s', w = w'} & = \prod_{i \in [\ell]} \Iv{s_i = s_i', w_i = w_i'}                                                                                                      \\
	                    & = \prod_{i \in [\ell]} \left(\Iv{s_i = w_i', w_i = s_i'} + \left( \Iv{s_i = s_i', w_i = w_i'} - \Iv{s_i = w_i', w_i = s_i'} \right) \right)             \\
	                    & = \sum_{S \subseteq [\ell]} \prod_{i \in S} \Iv{s_i = w_i', w_i = s_i'} \prod_{i \notin S} (\Iv{s_i = s_i', w_i = w_i'} - \Iv{s_i = w_i', w_i = s_i'}).
\end{align*}
We denote the product in the term $S \subseteq [\ell]$ as
\[
	K_S(s, w, s', w') := \prod_{i \in S} \Iv{s_i = w_i', w_i = s_i'} \prod_{i \notin S} (\Iv{s_i = s_i', w_i = w_i'} - \Iv{s_i = w_i', w_i = s_i'}).
\]
Thus, $\Iv{s = s', w = w'} = \sum_{S \subseteq [\ell]} K_S(s, w, s', w')$.
Intuitively, the set $S$ specifies for which indices of the strings we \emph{swap} $s_i'$ and $w_i'$; the factors in $[\ell] \setminus S$ are the correcting terms.

We define the \emph{anti-commutative indicator} to map characters $g, h, x, y \in [q]$ (for $g\neq h$) to
\[
	u_{gh}(x,y) := \begin{cases}
		1,  & (x,y)=(g,h),      \\
		-1, & (x,y)=(h,g),      \\
		0,  & \text{otherwise}.
	\end{cases}
\]

The anti-commutative indicator allows us to factor $K_S$ into two parts, such that one depends only on $s$ and $w$, and the other only on $s'$ and $w'$.
\begin{lemma}
	\label{lem:kx-using-swap-indicator}
	We have
	\begin{align*}
		K_S(s,w,s',w')=\sum_{\substack{g,h \in [q]^{\ell}                                \\ \forall i \notin S:  g_i < h_i}}\ \
		 & \prod_{i\in S}\Iv{s_i=g_i,w_i=h_i} \prod_{i\notin S}u_{g_i h_i}(s_i,w_i)      \\
		 & \prod_{i\in S}\Iv{s_i'=h_i,w_i'=g_i} \prod_{i\notin S}u_{g_i h_i}(s_i',w_i').
	\end{align*}
\end{lemma}
\begin{proof}
	We use two facts to rewrite each term in $K_S$.
	For characters $x, x', y, y' \in [q]$, we have
	\begin{equation}\label{eq:iv-conditioned}
		\Iv{x=y',y=x'}=\sum_{g,h\in[q]}\Iv{x=g,y=h}\Iv{x'=h,y'=g},
	\end{equation}
	and
	\begin{equation}\label{eq:iv-diff}
		\Iv{x=x',y=y'} - \Iv{x=y',y=x'} =\sum_{g<h}u_{gh}(x,y)u_{gh}(x',y').
	\end{equation}
    \Cref{eq:iv-conditioned} is immediate.
	For~\eqref{eq:iv-diff}, observe that if $x=y$, the sum vanishes and the indicators agree.
	If $x \ne y$, only $\{g,h\}=\{x,y\}$ contributes, and it gives exactly the left side.\footnote{Note that~\eqref{eq:iv-diff}  is the same as~\cite[Lemma~3.1]{Fischer26} with $U_{ij}=\Iv{(i,j)=(x,y)}$ and $V_{ij}=\Iv{(i,j)=(x',y')}$, which Fischer identifies as a form of the Cauchy--Binet formula.}

	Starting from the definition of $K_S$, we use~\eqref{eq:iv-conditioned} on each factor $i \in S$ and~\eqref{eq:iv-diff} on each factor $i \notin S$.
	This gives
	\[
		K_S(s, w, s', w') = \prod_{i \in S} \sum_{g_i, h_i \in [q]} \Iv{s_i = g_i, w_i = h_i} \Iv{s_i' = h_i, w_i' = g_i} \prod_{i \notin S} \sum_{g_i < h_i} u_{g_i h_i}(s_i, w_i) u_{g_i h_i}(s_i', w_i').
	\]
	Multiplying out this product gives a sum of terms, each of which corresponds to a choice of $g_i$ and $h_i$ for all $i \in [\ell]$.
	Grouping the factors of each term in the final sum into primed and unprimed variables yields the desired equality.
\end{proof}

Fix a set $S \subseteq [\ell]$.
To efficiently evaluate sums that contain a $K_S$ factor, we bound the support of~$K_S$.
For strings $s, w' \in [q]^\ell$, we let $\S_{s, w'} = \set{(w, s') \with K_S(s, w, s', w') \ne 0}$.

\begin{lemma}
	\label{lem:KS-nonzero-enumeration}
	We have $|\set{(s, w') \with |\S_{s, w'}| > 0}| \le q^{2\ell - |S|}$
	and $\sum_{s, w'} |\S_{s, w'}| \le q^{2\ell} 2^{\ell - |S|}$.
	Moreover, given~$s$ and $w'$, the elements of $\S_{s, w'}$ can be
	enumerated in $O(\ell(1 + |\S_{s, w'}|))$ time and $O(\ell)$ space.
\end{lemma}
\begin{proof}
	We count the ways to construct the strings $s, w, s', w' \in [q]^\ell$
	while maintaining that $K_S(s, w, s', w') \ne 0$.
	For $i \in S$, we have $s_i = w_i'$ and $w_i = s_i'$.
	The equality $s_i = w_i'$ at these indices yields that only
	$2\ell - |S|$ of the $2\ell$ characters of $s$ and $w'$ can be freely
	chosen, which establishes the first bound of the lemma.
	Among all four strings, if $i \in S$, there are $q^2$ ways to choose
	the characters $(s_i, w_i, s_i', w_i')$.
	For $i \notin S$, we consider two cases:
	\begin{itemize}
		\item
		      $s_i = w_i'$.
		      When choosing $w_i \ne s_i'$, both indicators in the
		      $i$-th factor of $K_S$ are $0$.
		      When choosing $w_i = s_i' = s_i$, both indicators are $1$
		      and cancel.
		      Thus we must set $w_i = s_i' \ne s_i$, in which case the
		      factor is $-1$.
		\item
		      $s_i \ne w_i'$.
		      Then the second indicator in the $i$-th factor of $K_S$
		      is $0$, so we must set $s_i = s_i'$ and $w_i = w_i'$,
		      in which case the factor is $1$.
	\end{itemize}
	Thus for each $i \notin S$, there is a binary decision for the case,
	and in each case there are at most $q^2$ possible choices of characters.
	Across all characters of all four strings, we get at most
	$(q^2)^{|S|} \cdot (2q^2)^{\ell - |S|}
	= q^{2\ell} \cdot 2^{\ell - |S|}$ choices, yielding the second bound.

	Finally, for the enumeration, given $s$ and $w'$, we first check
	whether they agree at the indices in $S$.
	If they do not, then $\S_{s, w'}$ is empty.
	Otherwise, for each $i \in S$, we choose $w_i = s_i'$ freely in $[q]$.
	For each $i \notin S$ with $s_i = w_i'$, we choose $w_i = s_i'$
	freely in $[q] \setminus \{s_i\}$.
	At every remaining index the pair $(w_i,s_i') = (w_i',s_i)$ is forced.
	By the cases above, the Cartesian product of these choices gives
	each element of $\S_{s, w'}$ exactly once.
	We enumerate this product using one counter per index.
	Since $q$ is constant, each counter has constant size.
	Initializing the counters takes $O(\ell)$ time, and advancing them
	and producing the next pair $(w,s')$ takes $O(\ell)$ time per pair.
	This gives $O(\ell(1 + |\S_{s, w'}|))$ total time and $O(\ell)$ bits
	of space.
\end{proof}

\section{Fast Clique Counting}
Fix $q=256$.
Without loss of generality, assume that $n = q^t$ for some $t \ge 2$ (if needed, we can insert up to $n \cdot q$ isolated vertices into $G$).
Identify vertices with $[q]^t$ and pairs of vertices with $[q]^\ell$, where $\ell=2t$.
The number of $6$-cliques in $G$ is $\frac{1}{6!} \sum_{a, b, c, d, e, f \in V(G)} \Iv{\set{a, b, c, d, e, f} \text{ is a $6$-clique}}$.
Our algorithm computes the same count as $Z/6!$ with
\[
	Z := \sum_{\substack{a,b,c,d,e,f\\ a',b',c',d' }} U(a,b,c,d,e)  \Iv{(a,b,c,d)=(a',b',c',d')}   P(a',b',c',d',e,f)
\]
where the sum runs over vertices of $G$ and we denote
\[
	\begin{aligned}
		U(a,b,c,d,e)       & = \Iv{\{a,b,c,d\} \text{ is a $4$-clique in } G \text{ and }ae,be \in E(G)}, \\
		P(a',b',c',d',e,f) & = P_1(a',c',e,f)\cdot P_2(b',d',e,f) \cdot \Iv{ef \in E(G)},              \\
		P_1(a',c',e,f)     & = \Iv{a'f,c'e,c'f \in E(G)},                                             \\
		P_2(b',d',e,f)     & = \Iv{b'f,d'e,d'f \in E(G)}.
	\end{aligned}
\]
Observe that when $\Iv{(a,b,c,d)=(a',b',c',d')} = 1$ in $Z$, the predicates $U$ and $P$ check that there is an edge between all pairs of vertices in $\set{a, b, c, d, e, f}$.
Among the variables in $\{a,b,c,d,a',b',c',d'\}$, $U$ uses only the unprimed variables $\set{a, b, c, d}$ whereas $P$ uses the primed variables $\set{a', b', c', d'}$.
The checks for the edges incident to $e$ and $f$ are distributed among $U$ and $P$.

\begin{figure}[tbp]
	\centering
	\begingroup
	\definecolor{utests}{HTML}{176A70}
	\definecolor{pone}{HTML}{B55B20}
	\definecolor{ptwo}{HTML}{704A9C}
	\definecolor{tipedge}{HTML}{555555}
	\definecolor{groupfirst}{HTML}{285A85}
	\definecolor{groupsecond}{HTML}{B55B20}

	{\small\textbf{(a) Clique decomposition used in the factored evaluation}}\par\smallskip
	\resizebox{\linewidth}{!}{%
		\begin{tikzpicture}[x=1cm,y=1cm,
				vertex/.style={
						circle,draw=black!65,fill=white,
						minimum size=7mm,inner sep=0pt,font=\normalsize
					},
				uedge/.style={draw=utests,line width=0.9pt},
				p one/.style={
						draw=pone,line width=0.9pt,
						dash pattern=on 3.5pt off 1.8pt
					},
				p two/.style={
						draw=ptwo,line width=0.9pt,
						dash pattern=on 4pt off 1.6pt on 0.9pt off 1.6pt
					},
				tip/.style={draw=tipedge,line width=0.9pt,densely dotted}
			]
			\path[use as bounding box] (0,1.55) rectangle (15.5,8.0);

			\coordinate (a)  at (3.4,6.55);
			\coordinate (b)  at (3.4,4.55);
			\coordinate (c)  at (1.2,6.55);
			\coordinate (d)  at (1.2,4.55);
			\coordinate (ap) at (13.7,6.55);
			\coordinate (bp) at (13.7,4.55);
			\coordinate (cp) at (9.3,6.55);
			\coordinate (dp) at (9.3,4.55);
			\coordinate (e)  at (6.15,7.05);
			\coordinate (f)  at (11.5,5.55);

			\foreach \x/\y in {a/b,a/c,a/d,b/c,b/d,c/d,a/e,b/e}
			\draw[uedge] (\x) -- (\y);

			\foreach \x/\y in {ap/f,cp/e,cp/f}
			\draw[p one] (\x) -- (\y);
			\foreach \x/\y in {bp/f,dp/e,dp/f}
			\draw[p two] (\x) -- (\y);

			\draw[tip] (e) to[out=-90,in=180] (7.35,3.65)
			-- (10.8,3.65) to[out=0,in=-90] (f);

			\foreach \id/\lab in {
			a/a,b/b,c/c,d/d,ap/{a'},bp/{b'},cp/{c'},dp/{d'},e/e,f/f
			}
			\node[vertex] at (\id) {$\lab$};

			\draw[uedge] (0.65,2.65) -- (1.45,2.65);
			\node[anchor=west,font=\small] at (1.6,2.65)
			{unprimed edges $U$};
			\draw[p one] (8.0,2.65) -- (8.8,2.65);
			\node[anchor=west,font=\small] at (8.95,2.65)
			{primed edges $P_1$};

			\draw[p two] (0.65,1.95) -- (1.45,1.95);
			\node[anchor=west,font=\small] at (1.6,1.95)
			{primed edges $P_2$};
			\draw[tip] (8.0,1.95) -- (8.8,1.95);
			\node[anchor=west,font=\small] at (8.95,1.95)
			{$ef$};
		\end{tikzpicture}%
	}

	\par\medskip
	{\small\textbf{(b) Clique decomposition used in the direct evaluation}}
	\par\smallskip
	\resizebox{\linewidth}{!}{%
		\begin{tikzpicture}[x=1cm,y=1cm,
				vertex/.style={
						circle,draw=black!65,fill=white,
						minimum size=7mm,inner sep=0pt,font=\normalsize
					},
				fixed/.style={vertex,fill=black!12,draw=black!75},
				inner first/.style={vertex,draw=groupfirst},
				inner second/.style={vertex,draw=groupsecond},
				first edge/.style={draw=groupfirst,line width=0.9pt},
				second edge/.style={
						draw=groupsecond,line width=0.9pt,
						dash pattern=on 3.5pt off 1.8pt
					}
			]
			\path[use as bounding box] (0,0.9) rectangle (15.5,8.0);

			\coordinate (a)  at (3.4,6.55);
			\coordinate (b)  at (3.4,4.55);
			\coordinate (c)  at (1.2,6.55);
			\coordinate (d)  at (1.2,4.55);
			\coordinate (ap) at (13.7,6.55);
			\coordinate (bp) at (13.7,4.55);
			\coordinate (cp) at (9.3,6.55);
			\coordinate (dp) at (9.3,4.55);
			\coordinate (e)  at (6.15,7.05);
			\coordinate (f)  at (11.5,5.55);

			\foreach \x/\y in {a/b,a/c,a/d,b/c,b/d,c/d,ap/f,bp/f}
			\draw[first edge] (\x) -- (\y);

			\foreach \x/\y in {a/e,b/e,cp/e,dp/e,cp/f,dp/f}
			\draw[second edge] (\x) -- (\y);
			\draw[second edge] (e) to[out=-90,in=180] (7.35,3.65)
			-- (10.8,3.65) to[out=0,in=-90] (f);

			\foreach \id/\lab in {a/a,b/b,cp/{c'},dp/{d'},f/f}
			\node[fixed] at (\id) {$\lab$};

			\foreach \id/\lab in {c/c,d/d,ap/{a'},bp/{b'}}
			\node[inner first] at (\id) {$\lab$};

			\node[inner second] at (e) {$e$};

			\draw[first edge] (0.65,2.65) -- (1.45,2.65);
			\node[anchor=west,font=\small] at (1.6,2.65)
			{first parenthesis of~\eqref{eqzdirect}};
			\draw[second edge] (8.0,2.65) -- (8.8,2.65);
			\node[anchor=west,font=\small] at (8.95,2.65)
			{second parenthesis of~\eqref{eqzdirect}};

			\node[circle,draw=black!75,fill=black!12,
				minimum size=4mm,inner sep=0pt] at (0.95,1.95) {};
			\node[anchor=west,font=\small] at (1.6,1.95)
			{fixed: $a,b,c',d',f$};

			\node[circle,draw=groupfirst,fill=white,
				minimum size=4mm,inner sep=0pt] at (0.95,1.25) {};
			\node[anchor=west,font=\small] at (1.6,1.25)
			{first sum: $c,d,a',b'$};

			\node[circle,draw=groupsecond,fill=white,
				minimum size=4mm,inner sep=0pt] at (8.4,1.25) {};
			\node[anchor=west,font=\small] at (8.95,1.25)
			{second sum: $e$};
		\end{tikzpicture}%
	}
	\endgroup

	\caption{Two decompositions of the fifteen edge tests.
		(a) The factors $U,P_1,P_2$ from Lemma~\ref{lem:factored-evaluation}, together with the separate
		edge $ef$.
		(b) In the direct evaluation~\eqref{eqzdirect}, the first parenthesis tests whether $\{a,b,c,d\}$ forms a $4$-clique, and whether $a'f,b'f \in E(G)$, while the second tests
		$ae,be,c'e,d'e,c'f,d'f,ef$.
		Shaded vertices are fixed for these two inner sums; the first
		sum ranges over $c,d,a',b'$, and the second over $e$.}
	\label{fig:edge-test-split}
\end{figure}
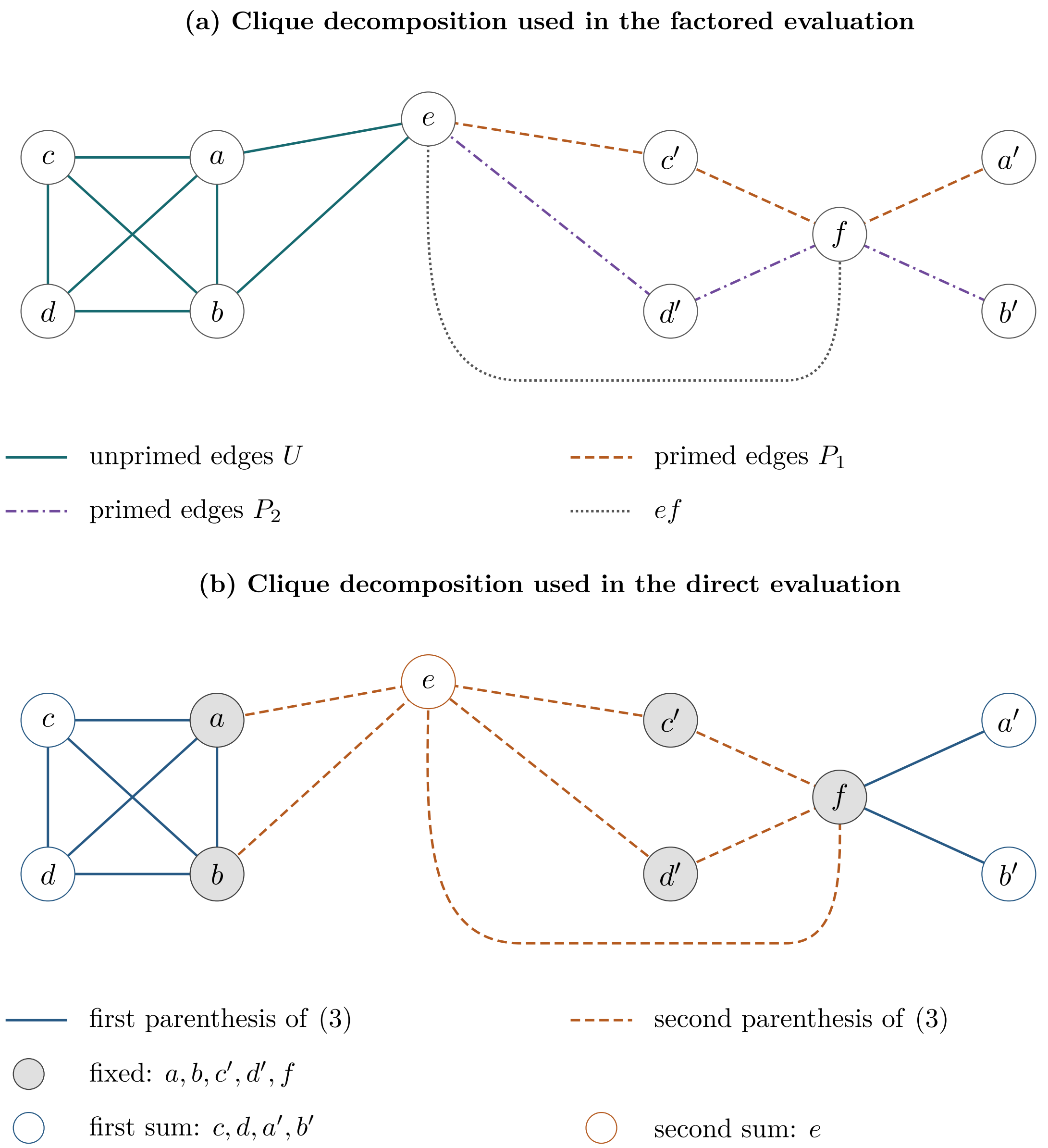

We split $Z$ into a family of similar sums by expanding the indicator.
Let $s,w,s',w' \in [q]^{\ell}$.
Recall that $\Iv{s=s',w=w'}=\sum_{S\subseteq[\ell]}K_S(s,w,s',w')$.
Going forward, we set $s = ab$, $s' = a'b'$, $w = cd$, and $w' = c'd'$.
For a fixed $S \subseteq [\ell]$, define
\[
	Z_S := \sum_{\substack{a,b,c,d,e,f\\ a',b',c',d' }} U(a,b,c,d,e)  K_S(ab,cd,a'b',c'd') P(a',b',c',d',e,f).
\]
We have $Z=\sum_{S\subseteq[\ell]}Z_S$.
It thus remains to show how we can compute $Z_S$ efficiently.
In the remainder of this section, we consider two ways to do so. The first exploits a factorization of $K_S$ whereas the second evaluates $K_S$ directly.
Both use $O(\log n)$ space, but have varying running time dependent on $|S|$.
\Cref{fig:edge-test-split} visualizes the two computations.

\begin{lemma}[Factored Evaluation]
	\label{lem:factored-evaluation}
	$Z_S$ can be evaluated in $\widetilde O\!\left(n^5+n^6 2^{|S|-t}\right)$ time.
\end{lemma}
\begin{proof}
	By using \cref{lem:kx-using-swap-indicator}, reordering the sums, and factoring the innermost sum, we obtain
	\begin{align*}
		Z_S = & \sum_{e}\sum_{\substack{g,h \in [q]^{\ell}                                                           \\ \forall i \notin S:  g_i < h_i}} \overbrace{\left( \sum_{a,b,c,d}U(a,b,c,d,e) \prod_{i\in S}\Iv{(ab)_i=g_i,(cd)_i=h_i}
		\prod_{i\notin S}u_{g_i h_i}((ab)_i,(cd)_i) \right)}^{L}                                                      \\
		      & \underbrace{\left( \sum_{a',b',c',d',f}P(a',b',c',d',e,f)\prod_{i\in S}\Iv{(a'b')_i=h_i,(c'd')_i=g_i}
			\prod_{i\notin S}u_{g_i h_i}((a'b')_i,(c'd')_i) \right)}_{R}.
	\end{align*}

	To compute $L$, we can enumerate all possibly-non-zero summands in $\widetilde{O}(2^{\ell-|S|})$ time.
	To see this, note that $(ab)_i=g_i$ and $(cd)_i=h_i$ are forced for $i \in S$, and for $i \notin S$, we have either that $(ab)_i=g_i$ and $(cd)_i=h_i$ or $(ab)_i=h_i$ and $(cd)_i=g_i$.
	We can enumerate all strings $ab$ and $cd$ with these properties in $O(\ell)$ time per string, and the value of each summand can be computed in $O(\ell)$ time.

    To compute $R$, we expand $P(a',b',c',d',e,f) = \Iv{ef \in E(G)}P_1(a',c',e,f)P_2(b',d',e,f)$ and simplify the factors based on single characters.\footnote{For example, for $i \in S$, the term $\Iv{(a'b')_i = h_i}$ turns into $\Iv{a'_i = h_i}$ if $i \le t$ or into $\Iv{b'_{i-t} = h_{ i}}$ if $i > t$.}
	We then factor the sum into two parts that depend only on $\set{a', c'}$ and $\set{b', d'}$ respectively.
	\[
		R = \sum_f \Iv{ef \in E(G)}
		\begin{aligned}[t]
			\cdot &
			\Biggl(
				\sum_{a',c'}P_1(a',c',e,f)
				\prod_{i\in S\cap[t]}\Iv{a_i'=h_i,c_i'=g_i}
				\prod_{i\in[t]\setminus S}u_{g_i h_i}(a_i',c_i')
				\Biggr)
			\\
			\cdot &
			\Biggl(
				\sum_{b',d'}P_2(b',d',e,f)
			\prod_{\substack{i\in[t] \\t+i\in S}} \Iv{b_i'=h_{t+i},d_i'=g_{t+i}}
			\prod_{\substack{i\in[t] \\t+i\notin S}} u_{g_{t+i}h_{t+i}}(b_i',d_i')
				\Biggr)
		\end{aligned}
	\]
	To evaluate $R$, we sum over $f$ and compute each summand by computing the two inner sums independently.
	Similar to the argument for computing $L$, both inner sums have at most $2^t$ possibly-non-zero terms and they can be enumerated efficiently:
	For the first, all non-zero summands satisfy $(a'_i,c'_i) \in \set{(g_i,h_i),(h_i,g_i)}$, and for the second all non-zero summands satisfy $(b'_i,d'_i) \in \set{(g_{t+i}, h_{t+i}), (h_{t+i},g_{t+i})}$.
	The value of each summand can be computed in $O(\ell)$ time.

    There are $q^{2|S|}\binom q2^{\ell-|S|}$ admissible choices of $(g,h)$.
	Hence, the total time to compute $Z_S$ is
	\[
		\widetilde O\!\left(n \cdot q^{2|S|}\binom q2^{\ell-|S|} \bigl(2^{\ell-|S|}+n2^t\bigr) \right)
		 \le \widetilde O\!\left( n \cdot q^{2\ell}2^{|S|-\ell} \bigl(2^{\ell-|S|}+n2^t\bigr) \right)
		 = \widetilde O\!\left(n^5+n^6 2^{|S|-t}\right)
	\]
	where we used $q^{2\ell}=n^4$ and that the $\widetilde O$ hides $\ell=O(\log n)$ factors.\footnote{
		Observe that the $2^{|S| - \ell}$ that comes from the denominator $2$ of $\binom q 2$ is the essential part in decreasing the $n^6$ summand.
		Limiting the sums to just $g_i < h_i$ thus is significant for the speedup.
	}
\end{proof}
\begin{lemma}[Direct Evaluation]
	\label{lem:direct-evaluation}
	$Z_S$ can be evaluated in $\widetilde O\left(n^5 2^{\ell-|S|}+n^6q^{-|S|}\right)$ time.
\end{lemma}
\begin{proof}
	For fixed $a,b,c',d',f$, all edge tests involving $e$ depend
	only on these fixed vertices and $e$. We can therefore write $Z_S$ as
	\begin{equation}\label{eqzdirect}
		\begin{aligned}
			Z_S=\sum_{a,b,c',d',f}
			 & \Biggl(
			\sum_{c,d,a',b'}K_S(ab,cd,a'b',c'd')\cdot
			\Iv{\set{a,b,c,d}\text{ is a }4\text{-clique in }G}
			\Iv{a'f,b'f\in E(G)}
			\Biggr)    \\
			\cdot
			 & \Biggl(
			\sum_e\Iv{ae,be,c'e,d'e,c'f,d'f,ef\in E(G)}
			\Biggr).
		\end{aligned}
	\end{equation}

	We evaluate~\eqref{eqzdirect} by exhaustively iterating over $a$, $b$, $c'$, $d'$, and $f$.
	We then use \cref{lem:KS-nonzero-enumeration} to enumerate the values of $a'$, $b'$, $c$, and $d$ for which the $K_S$ factor is non-zero and compute the first inner sum.
	If this sum is non-zero, we compute the second inner sum over $e$ by exhaustive iteration.

    For each fixed $f$, the pairs $(s,w')=(ab,c'd')$ range over $[q]^\ell \times [q]^\ell$.
    By \cref{lem:KS-nonzero-enumeration}, and since each enumerated term takes $O(\ell)$ time to evaluate, the total time to compute all first inner sums is
    \[
    	O\!\left(\ell n^5 + \ell n \sum_{s,w'} |\S_{s,w'}|\right)
    	= O\!\left(\ell n^5 + \ell n q^{2\ell} 2^{\ell-|S|}\right)
    	= \widetilde O\!\left(n^5 2^{\ell-|S|}\right),
    \]
    where we used $q^{2\ell}=n^4$.
    The first inner sum can be non-zero only if $\S_{s, w'}$ is non-empty.
	Thus, by \cref{lem:KS-nonzero-enumeration}, the total time to compute all second inner sums is $\widetilde O(n \cdot q^{2\ell - |S|} \cdot n)$.
	Adding the two costs and using $q^{2\ell}=n^4$, the time this evaluation takes is $ \widetilde O(n^5 2^{\ell-|S|}+n^6q^{-|S|}) $.
\end{proof}
We are now ready to prove our main theorem:
\begin{proof}[Proof of \cref{thm:clique}]
	To evaluate $Z$, we iterate over all sets $S$ and compute $Z_S$ using whichever of the two methods from \cref{lem:factored-evaluation,lem:direct-evaluation} is faster.
	Both first terms are at most $n^5 2^{\ell-|S|}$.
	By the binomial theorem, $\sum_{S\subseteq[\ell]}2^{\ell-|S|} =\sum_{j=0}^{\ell}\binom{\ell}{j}2^{\ell-j}=3^{\ell}=9^t$.
	Therefore the total time is
	\begin{equation}\label{eq:totaltime}
		\widetilde O\!\left(n^5 9^{t}+n^6\sum_{S\subseteq[\ell]}\min\set*{2^{|S|-t},q^{-|S|}}\right).
	\end{equation}
    Using the inquality $\min\{x,y\}\leq\sqrt{xy}$ of $x,y \in \mathbb{R}_{\geq 0}$ and the binomial theorem, we obtain
\[
\begin{aligned}
\sum_{S\subseteq[\ell]}
    \min\{2^{|S|-t},q^{-|S|}\}&\leq 2^{-t/2}\sum_{S\subseteq[\ell]}(2/q)^{|S|/2}\\
&=2^{-t/2}(1+\sqrt{2/q})^{2t} =2^{-t/2}\left(1+\sqrt{1/128}\right)^{2t}
<2^{-t/4},
\end{aligned}
\]
where we use that
\[
\frac{(1+\sqrt{1/128})^2}{\sqrt{2}}
<0.838<2^{-1/4}.
\]
Since $n=256^t$, we have $9^t<\sqrt{n}$ and
$2^{-t/4}=n^{-1/32}$.
Substituting into~\eqref{eq:totaltime} gives
\[
\widetilde O\!\left(n^{11/2}+n^{6-1/32}\right)
=O(n^{5.97}),
\]
where $6-1/32=5.96875<5.97$ leaves enough slack to absorb
the polylogarithmic factors.

	To see that this algorithm uses only $O(\log n)$ space, note that there is only a constant number of sums and that each iterates over $n^{O(1)}$ values.
	At any point, only a constant number of variables and intermediate results is stored.
    When enumerating non-zero terms of a sum, we use counters of size $O(\ell) = O(\log n)$.
	All intermediate results are of absolute value $n^{O(1)}$, thus taking $O(\log n)$ bits.
\end{proof}

\section{Open Questions}
Our result implies that \textsc{$k$-Clique} can be solved in $O(n^{k-0.03})$ time and $O(\log n)$ space for fixed~$k \ge 6$. Can \textsc{$k$-Clique} be solved in $O(n^{k-\varepsilon})$ time and $\log^{O(1)}(n)$ space for some $\varepsilon >0$ and $k = 3,4,5$?
Another open question in the light of the recent work~\cite{ALPPW26,CIMY26,Fischer26} is whether $k$-hyperclique in $3$-uniform hypergraphs can be solved in time $O(n^{k-\varepsilon})$ for some $\varepsilon>0$ and $k$ and $\log^{O(1)}(n)$ space.

\section*{AI Methodology}
The main construction of this paper was found by OpenAI's GPT-6 Astra, after we asked it whether sparse factorizations of the $k$-clique tensor can be extracted from~\cite{VWZ26} in order to derive an $O(n^{k-\varepsilon})$ time and $\log^{O(1)}(n)$ space algorithm for \textsc{$k$-Clique} for some $\varepsilon \in \mathbb{R}_{>0}$ and $k \in \mathbb{Z}_{>0}$. Afterwards ChatGPT was also used for the write-up. The authors reorganized and rewrote the proof.

\tiny
\printbibliography

\end{document}